\input harvmac
\overfullrule=0pt
\def\Title#1#2{\rightline{#1}\ifx\answ\bigans\nopagenumbers\pageno0\vskip1in
\else\pageno1\vskip.8in\fi \centerline{\titlefont #2}\vskip .5in}

%
\Title{\vbox{\baselineskip12pt
\hbox{hep-th/9712251}\hbox{HUTP-97/A106}}}
{\vbox{\centerline {Black Hole Entropy from }
\centerline {Near--Horizon Microstates } }}


\centerline{Andrew Strominger}

\bigskip\centerline{Jefferson Laboratory of Physics}
\centerline{Harvard University}\centerline{Cambridge, MA 02138}
\def\ads {AdS$_3$}
\def\slr{$SL(2,R)_L\otimes SL(2,R)_R$}


\vskip .3in

\centerline{\bf Abstract}

\smallskip
\noindent Black holes whose near--horizon geometries are locally, but not
necessarily globally, \ads\ (three--dimensional anti-de Sitter space) are
considered. Using the fact that 
quantum gravity on \ads\ is a conformal field theory, we microscopically 
compute the black hole entropy from the asymptotic growth of states.
Precise numerical agreement with the Bekenstein-Hawking area formula for 
the entropy is found.
The result pertains to any consistent quantum theory of gravity, and does
not use string theory or supersymmetry.

\Date{}

\lref\jbmh{J.\ D.\ Brown and M.\
Henneaux, {\sl Comm. Math. Phys.} {\bf 104} (1986) 207.}
\lref\at{A.\ 
Achucarro and P.\ Townsend, 
{\sl Phys. Lett.} {\bf B180} (1986).}
\lref\ew{E.\ Witten, {\sl
Nucl. Phys.} {\bf B311} (1989) 46.} 
\lref\clp{S.\ Carlip, Phys. Rev. {\bf D51} (1995) 632.}
\lref\ascv{A.\ Strominger and C.\ Vafa, {\sl Phys. Lett.} {\bf B379}
(1996) 99.}
\lref\dps{M.\ Douglas, J.\
Polchinski and A.\ Strominger, hep-th/9703031.}
\lref\jm{J.\ Maldacena,
hep-th/9711200.}
\lref\sh{S.\ Hyun, hep-th/9704005.}
\lref\bph{H.\ Boonstra, B.\ Peeters, and K.\ Skenderis, hep-th/9706192.}
\lref\ss{K.\ Sfetsos and K.\ Skenderis, hep-th/9711138.}
\lref\btz{M.\ Banados, C.\ Teitelboim and J.\ Zanelli, {\sl
Phys. Rev. Lett.} {\bf 69} (1992) 1849.}
\lref\bhtz{M.\ Banados, M.\ Henneaux, C.\ Teitelboim, and J.\ Zanelli, {\sl
Phys.Rev.} {\bf D48} (1993) 1506.}
\lref\chv{O.\ Coussaert, M.\ Henneaux and P.\ van Driel, Class. 
Quant. Grav {\bf 12} (1995) 2961.}
\lref\ch{O.\ Coussaert and M.\ Henneaux, Phys. Rev. Lett {\bf 72} (1994) 183.}
\lref\chtwo{O.\ Coussaert and M.\ Henneaux, hep-th/9407181.}
\lref\nha{J. W. York, Phys. Rev. {\bf D28} (1983) 2929.}
\lref\rez{B. Reznik, Phys. Rev. {\bf D51} (1995) 1728.}
\lref\nhb{W. H. Zurek and K. S. Thorne, Phys. Rev. Lett. {\bf 54} (1985)
          2171.}
\lref\nhc{ J. A. Wheeler, {\it A Journey into Gravity and Spacetime}
Freeman, N.Y. (1990).}
\lref\nhd{ G. 't Hooft, Nucl. Phys. {\bf B335} (1990) 138.}
\lref\nhe{L. Susskind, L. Thorlacius and R. Uglum, 
Phys. Rev. {\bf D48} (1993) 3743.}
\lref\nhf{V. Frolov and I. Novikov, Phys. Rev. {\bf D48} (1993) 4545.}
\lref\nhg{M. Cvetic and A. Tseytlin, Phys. Rev. {\bf D53} (1996) 5619.}
\lref\nhh{F. Larsen and F. Wilczek, Phys. Lett. {\bf B375} (1996) 37.}
\lref\hw{G. T. Horowitz and D. Welch, Phys. Rev. Lett. {\bf 71} (1993) 328.}
\lref\gth{G. 't Hooft, gr-qc 9310026.}
\lref\lss{L. Susskind, J. Math, Phys. {\bf 36} (1995) 6377.}
\lref\ls{D. A . Lowe and A. Strominger, Phys. Rev. Lett {\bf 73} (1994) 1468.}
\lref\jc{J. A. Cardy, Nucl. Phys. {\bf B270} (1986) 186.}
\lref\scct{S. Carlip and C. Teitelboim, Phys. Rev. {\bf D51} (1995) 622. }

\newsec{Introduction}

The idea that the black hole entropy should be accounted for by microstates
near the black hole horizon has great appeal and a long history \refs{
\nha \nhb \nhc \nhd \nhe \nhf \clp \nhg -\nhh}. One reason for this 
is that a demonstration that 
the horizon has of order one degree of freedom per Planck 
area would provide a statistical explanation of the area formula 
for the entropy. Such a picture might also shed light on the 
information puzzle, in that the information is more safely stored on the
surface than in the causally inaccessible black hole interior.

While much has been learned, 
attempts at a precise statistical accounting for the entropy
along these lines have been hampered both by the
ultraviolet problems in quantum gravity as well as the infinite number of
low-energy modes which are nevertheless high-frequency because of the
large near--horizon redshifts\foot{With the notable exception of the
topological theory of gravity considered in 
\clp\ and discussed below.}. In practice it has not been clear which
modes should or should not be counted, and the results appear cutoff
dependent.

Recently a statistical derivation of the black hole entropy has been given
for some cases in string theory \ascv. This derivation employed a 
continuation to a weak-coupling region where the
black hole is treated as a pointlike object and its microstates are counted
by a certain conformal field theory. This construction did not directly address the
issue of whether or not, 
in regimes for which the black hole is not 
effectively pointlike and has a clear
horizon, the microstates are in any sense near the horizon. 

In this paper we will address this issue in the context of black holes
whose near--horizon geometry is locally \ads. This includes
many of the string theory examples and the
three--dimensional BTZ black hole \btz. We will microscopically derive the black
hole entropy by counting excitations of \ads. Our derivation relies only on
general properties of a diffeomorphism-invariant theory and will not use
string theory or supersymmetry.

Our basic result follows quickly from prior results in the literature.
Some time ago, Brown and Henneaux \jbmh\ made the remarkable observation
that the asymptotic symmetry group of \ads\ is generated by 
(two copies of) the Virasoro
algebra, and that therefore {\sl any consistent quantum theory of gravity
on \ads\ is a conformal field theory}\foot{Maldacena has recently shown,
from a totally different perspective, 
that certain string compactifications to \ads\ are conformal field
theories \jm.}. They further computed the value of
the central charge as $c={3 \over 2G\sqrt{-\Lambda}}$, where $G$ is
Newton's constant and $\Lambda$ is the cosmological constant. In this paper we
simply apply Cardy's 
formula \jc\ for the asymptotic growth of states for a conformal
field theory of central charge $c$ to microscopically compute the black
hole entropy. Precise agreement with the Bekenstein-Hawking area formula is
found.

The conformal field theory that describes the black hole microstates 
lives on a ($1+1$)--dimensional cylinder 
surrounding the black hole. Since all the information is 
encoded in this cylinder and remains outside the horizon 
there is no information loss in this picture.  

This paper is organized as follows: In Section 2 the result of
Brown and Henneaux is reviewed. In Section 3 the BTZ black hole
is reviewed. 
In Section 4 other relevant black holes which approach \ads\ near 
the horizon are discussed. In Section 5 the entropy is microscopically 
computed. In Section 6 we
relate the present work to other derivations of the black hole entropy as
well as Maldacena's recent work on near-horizon dynamics in string theory 
\jm. We close with discussion in Section 7.

\newsec{Quantum Gravity on AdS$_3$ as a Conformal Field Theory}

In this section we review the results of \jbmh. Three-dimensional
gravity coupled to matter is described by the action  \eqn\ctn{S= {1\over 16
\pi G} \int d^3 x \sqrt{-g} (R + {2\over \ell^2}) + S_m,} where $S_m$ is
the matter action, the cosmological constant is $\Lambda = -{1\over \ell^2}$
and we have omitted surface terms. The matter action will not play an
important role in the following, but we include it here to stress the
generality of our considerations.
In the following we will be interested in a semiclassical description,
which requires that the cosmological constant is small in Planck 
units, or equivalently
\eqn\lcs{\ell \gg G.}

\ctn\ has the \ads\
vacuum solution 
\eqn\adst{ds^2= -\left({r^2\over \ell^2} + 1 \right) dt^2 +
\left( {r^2\over \ell^2 }+ 1 \right)^{-1} dr^2 + r^2 d\phi^2,} 
where $\phi$
has period $2\pi$. \ads\ is the $SL(2,R)$ group manifold and accordingly
has an \slr\ isometry group. In order to
define the quantum theory on \ads, we must specify boundary conditions at
infinity. These should be relaxed enough to allow finite mass excitations
and the action of \slr, but tight enough to allow a well-defined
action of the diffeomorphism group. 
This requires \jbmh\foot{Matter fields, if present, are assumed to fall off
rapidly enough so as not to affect the asymptotic form of the 
symmetry generators.}
\eqn\asb{\eqalign{g_{tt}&=
{-r^2\over \ell^2} + {\cal O} (1),\cr g_{t\phi}&=
{\cal O}(1),\cr
g_{tr}&= {\cal O} ({1 \over r^3}),\cr
g_{rr}&= {\ell^2 \over r^2} + {\cal O} ({1 \over r^4}),\cr g_{r\phi}&={\cal
O} ({1 \over r^3}),\cr g_{\phi\phi}&= r^2 + {\cal O} (1).\cr}} Allowed
diffeomorphisms are generated by vector fields $\zeta^a (r, t,\phi)$ which
preserves \asb. These are of the form 
\eqn\ctm{\eqalign{\zeta^t &=\ell (T^+
+ T^-) + {\ell^3\over 2r^2} (\partial_+^2 T^+ + \partial_-^2 T^-) + {\cal O}
({1\over r^4}),\cr \zeta^\phi &=T^+ - T^- -{\ell^2 \over 2r^2}
(\partial_+^2 T^+ - \partial_-^2 T^-)  + {\cal O} ({1 \over r^4}),\cr
\zeta^r &=-r(\partial_+ T^+ + \partial_- T^-) + {\cal O} ({1\over r}),\cr}}
where $2\partial_\pm \equiv \ell{\partial\over \partial t} \pm
{\partial\over \partial \phi}$ and preservation of \asb\ requires that 
$T^\pm$ depend on $r,\; \phi$ and $t$ only as $T^\pm(r, t,\phi) = T^\pm ({t \over \ell} \pm
\phi)$ so that $\partial_\pm T^\mp =0$.

Diffeomorphisms with $T^\pm =0$ fall off rapidly at infinity and should be
considered ``pure gauge transformations". In the quantum theory the
corresponding generators will annihilate physical states. The
diffeomorphisms with nonzero $T_\pm$ modulo the pure gauge transformations
comprise the asymptotic symmetry group. Let us denote the 
generators of these
diffeomorphisms by 
$$L_n,~~~~~~~~ \; \bar L_n \;\;\;\;\;\; -\infty < n < \infty~,$$
where $L_n \; (\bar L_n)$ generates the diffeomorphism with 
$T^+ = e^{in({t\over \ell} +
\phi)}$ $\left( T^- = e^{in({t\over \ell} -\phi)} \right)$. The 
generators obey the algebra 
\eqn\vlg{\eqalign{ [L_m,L_n] &= (m-n) L_{m+n} + {c\over 12} (m^3 -m)
\delta_{m+n,0},\cr
[\bar L_m, \bar L_n] &=(m-n) \bar L_{m+n} + {c\over 12} (m^3 -m)
\delta_{m+n,0},\cr
[L_m,\bar L_n] &=0,\cr}}
with
\eqn\cch{c={3\ell \over 2G}.}
The explicit computation of the central charge $c$ proceeds \jbmh\ by taking the matrix
element of \vlg\ in the \ads\ vacuum \adst. The central charge is then
related to an integral of the vector field parameterizing the
diffeomorphism generated by $L_m$ over the vacuum \adst\ perturbed by
$L_{-m}$. We note that in the semiclassical regime \lcs\
\eqn\scg{c\gg 1.}

\vlg\ is of course the Virasoro algebra. Since the physical states of
quantum gravity on \ads\ must form a representation of this algebra, we
have the remarkable result \jbmh\ {\sl quantum gravity on \ads\ is a
conformal field theory with central charge $c={3\ell \over 2G}.$} The
conformal field theory lives on the $(t,\phi)$ cylinder at spatial
infinity.

\newsec{The BTZ Black Hole}
An important example to which our considerations apply is the 
BTZ black hole \refs{\btz, \bhtz}. In this section we recall a 
few of its salient features. Further details can be found in \bhtz. 

The metric for a black hole of mass $M$ and angular momentum $J$ 
is 
\eqn\btzm{ds^2=-N^2dt^2+\rho^2(N^\phi dt +d\phi)^2 +{r^2 \over
N^2\rho^2}dr^2,}
with 
\eqn\dfs{\eqalign{N^2&={r^2(r^2-r_+^2) \over \ell^2 \rho^2},\cr
                 N^\phi &=-{4GJ \over  \rho^2},\cr
                 \rho^2&=r^2+4G M\ell^2-\half r^2_+,\cr
                  r_+^2&=8G\ell\sqrt{M^2\ell^2-J^2},}}
where $\phi $ has period $2\pi$.
These metrics obey the boundary conditions \asb, and the black holes are
therefore  in the Hilbert space of the conformal field theory. The 
Bekenstein-Hawking black hole entropy is 
\eqn\sen{S={{\rm Area} \over 4G}={\pi\sqrt{16GM\ell^2+2r_+^2}\over 4G}.}

It is convenient to choose the additive constants in $L_0$ and $\bar L_0$ 
so that they vanish for the $M=J=0$ black hole. One then has
\eqn\mab{M= {1 \over \ell} (L_0 + \bar L_0),}
while the angular momentum is
\eqn\jab{J= L_0 - \bar L_0.}
The metric for the $M=J=0$ black hole is
\eqn\zmb{ds^2=-{r^2 \over \ell^2}dt^2+r^2d\phi^2+{\ell^2 \over r^2}dr^2.}
This is $not$ the same as \ads\ metric \adst\ which has negative
mass $M=-{1 \over 8G}$.
Locally they are equivalent since there is locally only one constant 
curvature metric in three dimensions. However they are inequivalent 
globally. The family of metrics \btzm\ can be obtained from \adst\ 
by a discrete identification 
followed by a 
coordinate transformation. Hence the black holes are topologically 
inequivalent to \ads. If one attempts to deform the black hole solution to
\ads\ by varying $M$, naked singularities are encountered between
$M=0$ and $M=-{1 \over 8G}$. We will view the nonzero mass black holes as
excitations of the vacuum described by the $M=0$ black hole.

This raises the question of the role played by the \ads\ vacuum. 
A beautiful answer to this question was given in \ch\ which however 
requires supersymmetry. In a supersymmetric conformal field 
theory the Ramond 
ground state, which has periodic boundary conditions and is annihilated by
the supercharge, has $M=0$. We identify this with the zero-mass black 
hole \zmb. The Neveu-Schwarz ground state has
antiperiodic boundary conditions and is not supersymmetric. It 
has a mass shift 
\eqn\rft{L_0=\bar L_0 =-{c \over 24}.}
Using the formula \cch\ for $c$ and \mab\ for $M$ we find
the energy of the  Neveu-Schwarz ground state 
is 
\eqn\nsg{M=-{1 \over 8G},}
which leads us to identify it with \ads. Further evidence for this
identification comes from the fact that the covariantly constant
\ads\ spinors are antiperiodic 
under $\phi \to \phi +2\pi$ (because it is a $2\pi$ rotation), 
while the covariantly constant spinors in the extremal $\ell M=J$ 
black hole geometries are periodic \ch. Additionally it can be seen 
in euclidean space \scct\ that 
the coordinate transformation relating \adst\ to \zmb\ is exponential 
in $t$ and $\phi$, just like the map from the plane to the cylinder. 

In a theory with local dynamics, a non-extremal BTZ black hole 
in empty space will Hawking radiate. However unlike the higher dimensional 
examples, there is (with appropriate boundary conditions at infinity) 
a stable, non-extremal endpoint corresponding to 
a black hole in thermal equilibrium with a radiation bath \rez. 
This is possible because an infinite radiation bath in \ads\ has finite 
energy due to an infinite temperature redshift at infinity. The 
generic hamiltonian eigenstate of the \ads\ conformal field theory 
presumably corresponds to such an equilibrium state. 

\newsec{Other Examples}
 The Bekenstein-Hawking entropy of a black hole depends only on the 
area of its horizon. In order to understand it only the near-horizon 
geometry of the black hole is relevant. Hence the considerations of 
this paper apply to any black hole whose near horizon geometry is 
\ads\  up to global identifications. There are many examples of this. 

One example, considered in the string theory context in 
\ascv, is black strings in six
dimension with charges $Q_1$, $Q_5$ and longitudinal momentum 
$n$. The near horizon geometry of this black string is 
locally 
$AdS_3 \times S^3 \times M^4$ with $M^4$ either $K^3$ or $T^4$ \sh. This may be
regarded as quantum gravity on \ads\ with an infinite tower of matter fields
from both massive string states and Kaluza-Klein modes of the $S^3 \times
M^4$ compactification. Therefore the states are a representation of the
Virasoro algebra \vlg.

The longitudinal direction along this black string lies within the \ads.
The black string can be periodically identified to give a black hole 
in five dimensions. The near-horizon geometry is then a BTZ 
black hole. For $n=0$ one obtains the $M=J=0$ black hole, 
while the generic black string yields the generic BTZ black hole. 

An example involving four-dimensional extremal charged black holes is 
considered in \ls, where the $U(1)$ arises from an internal circle. 
The near horizon geometry involves a $U(1)$ bundle over AdS$_2$. 
The geometry of the bundle is a quotient of \ads. The 
discrete identification group 
lies entirely within one of the two $SL(2,R)$s. Such geometries are
considered in \chtwo.

\newsec{Microscopic Derivation of the Black Hole Entropy}

We wish to count the number of excitations of the \ads\ vacuum with mass $M$
and angular momentum $J$ in the semiclassical regime of  
large $M$. According to \mab\ and \cch\ large $M$ implies 
\eqn\nrl{n_R +n_L\gg c,~~}
where $n_R$ ($n_L$) is the eigenvalue of $L_0$ ($\bar L_0$).
The asymptotic growth of the number states of
a conformal field theory with central charge $c$ is then
given by \jc\ 
\eqn\ssqrts{S = 2\pi \sqrt{{cn_R \over 6}} + 2 \pi \sqrt{{c n_L \over 6}}.}
Using \cch\ , \mab\ and \jab\ , this is
\eqn\lsqrts{S= {\pi} \sqrt{ \ell  (\ell M + J)\over 2G} + 
                       {\pi}\sqrt{ \ell  (\ell M - J)\over 2G},}
in exact agreement with the Bekenstein-Hawking result \sen\
for the BTZ black hole.

\newsec{Relation to Previous Derivations}

The microscopic derivation of the previous section rests on a key
assumption: the required $2+1$ quantum theory of gravity must exist. The
details of the theory are not important, except in that the states must as
discussed behave properly under diffeomorphisms. These are several
constructions of such theories in which microscopic derivations of the
entropy have previously been given. It is instructive to compare these with
the present derivation.

The first is Carlip's derivation \clp\ of the entropy in pure $2+1$ gravity
with no matter. This theory has no local degrees of freedom, and can be
recast as a topological Chern-Simons theory \refs{\at,\ew}. There are
nevertheless boundary degrees of freedom at the black hole horizon which
are described by a $1+1$ conformal field theory. Enumeration of the 
boundary states
yields the Bekenstein-Hawking entropy. This 
boundary conformal field theory has a different central charge 
($c \sim 6$ for large black holes) and so is not the same as 
the conformal field theory 
discussed here. Nevertheless it seems likely there is some 
connection between the approaches, which needs to be better 
understood. Relevant work in this direction 
appeared in \chv\ where it was shown that, for a boundary at 
infinity rather than the horizon, the Chern-Simons theory could 
be recast as a Liouville conformal field theory living on the boundary. 
An adaptation to string theory was discussed in \ss.

The second example is the five dimensional 
string theory black hole with charges $Q_1$, $Q_5$ and $n$. It was
indeed found in \ascv\
that the black hole states are described by a $c=6Q_1
Q_5$ conformal field theory. This agrees with \cch\ since in the
effective three-dimensional theory $\ell = 2\pi\alpha^\prime
\sqrt{g} (Q_1 Q_5)^{1/4}/V^{1/4} $ and $G^{-1} = 2V^{1/4}
(Q_1
Q_5)^{3/4}/\pi\alpha^\prime\sqrt{g}$.\foot{$V$ is the four-volume 
associated with the
compactification to six dimensions and we use units in which the 
ten-dimensional Newton's constant is given 
by $G=g^2 8\pi^6(\alpha^\prime)^4$.}

What is the relation between these two derivations of the 
black hole conformal field 
theory?
The string theory black hole has two
descriptions: one as a (string--corrected) supergravity solution, and the
other as a bound state of $Q_1$ D-onebranes and $Q_5$ D-fivebranes 
with momentum $n$. For
large $gQ$, but small string coupling $g$, string perturbation theory about
the supergravity solution generally provides a good description. For small
$gQ$, D-brane perturbation theory is generally good. It might 
appear there is no overlap in the regions of validity 
of the two descriptions. However, D-brane and
supergravity descriptions have an overlapping region of validity for large
$gQ$ in the near--horizon small $r$ region \dps. The supergravity picture is
good because the horizon is a smooth place even though $r$ is small. The
D-brane picture is also useful because small $r$ means low energies and
higher dimension corrections to the D-brane gauge theory are suppressed.
However, since $gQ$ is large, it is a large-$N$ gauge theory in this region.

This observation 
was made more precise in \jm\  
(along with interesting generalizations to AdS$_n$)with the introduction of a
certain scaling limit (see also \refs{\sh, \bph, \ss}). In this limit, on the
supergravity side, only the near--horizon theory of strings on \ads\
remains, while on the D-brane side one has only the conformal field theory
limit of the D-brane gauge theory.
It was further noted \jm\ that the equivalence of these two theories was
consistent with the global \slr\ symmetry of the ground state. With
the \ads\ boundary conditions defined in \jbmh, one can go a step further and
relate the action of the full  $local$ conformal group on both sides.

Of course, string theory enables one to go well 
beyond the considerations of this
paper. For example one 
can not only determine the central charge of the theory (which
is all that is needed for the entropy), but the exact conformal theory and 
degeneracies at every mass level.
\newsec{Discussion}
Where exactly are the states accounting for the black hole entropy? 
The states of the \ads\ conformal field theory are associated to the 
$(t, \phi)$ cylinder, and not with any particular value of the radius 
$r$. The dynamics of the theory can be described in terms of evolution 
on this cylinder without introducing fields which depend on $r$. 
In this description nothing crosses the horizon, and there is no
information loss. 

The disappearance of $r$ in the description of the 
quantum theory is not surprising in the context of pure $2+1$ 
gravity, 
since that is a topological theory with 
only boundary degrees of freedom. However our discussion also applies to
string theory in which one might expect new degrees of freedom at every
value of $r$. This is a concrete example of the holographic principle 
advocated in \refs{\gth,\lss}.

It would certainly be of interest to understand in greater detail the 
nature of the quantum states and dynamics in explicit examples. 
Consider the string theory black hole
made from the compactification of 
$Q_5$ NS fivebranes and $Q_1$ fundamental strings. In this 
case the near-horizon string theory can be represented by a 
semi-conventional $worldsheet$ conformal field theory:
a level $Q_5$ $SL(2,R)$ WZW model, with discrete identifications 
from compactification \refs{\hw,\ls,\nhg}.  The spectrum of this 
theory is not well-understood because it is noncompact. However perhaps
it can be better organized utilizing the action of the full conformal 
group on the  $SL(2,R)$=\ads\ target space. The Hilbert space for 
pure $2+1$ gravity also needs to be better understood, perhaps using the 
representation as a Liouville theory \chv.

\centerline{\bf Acknowledgements}
I would like to thank J. Maldacena for stimulating conversations. 
This work was supported in part by DOE grant DE-FG02-96ER40559.

\listrefs

\bye